\documentclass[aps,prb,twocolumn,groupedaddress]{revtex4}

\usepackage{graphicx,amssymb}

\begin{document}

\title{Coexistence of a triplet nodal order-parameter and a singlet order-parameter at the interfaces of ferromagnet-superconductor Co/CoO/In junctions}
\author{ S Hacohen-Gourgy}
\email{hacohe@post.tau.ac.il}
\author{B Almog}
\author{G Deutscher}
\affiliation{Raymond and Beverly Sackler School of Physics and Astronomy, Tel-Aviv University, 69978 Tel-Aviv, Israel}
\date{\today}

\begin{abstract}
We present differential conductance measurements of Cobalt / Cobalt-Oxide / Indium planar junctions, 500nm x 500nm in size. The junctions span a wide range of barriers, from very low to a tunnel barrier. The characteristic conductance of all the junctions show a V-shape structure at low bias instead of the U-shape characteristic of a s-wave order parameter. The bias of the conductance peaks is, for all junctions, larger than the gap of indium. Both properties exclude pure s-wave pairing. The data is well fitted by a model that assumes the coexistence of s-wave singlet and equal spin p-wave triplet fluids. We find that the values of the s-wave and p-wave gaps follow the BCS temperature dependance and that the amplitude of the s-wave fluid increases with the barrier strength.

\end{abstract}

\maketitle

\section{Introduction}
When a superconductor is brought in contact with a metal Cooper pairs penetrate into the normal side causing the Proximity effect\cite{ParksProximity}. If the metal is a ferromagnet the exchange field tends to align the spins. The Cooper pairs in a singlet state, penetrating the ferromagnet, are then destroyed at the interface. But, under certain circumstances, such as spin scattering or inhomogeneous magnetization, the exchange field can induce a triplet state\cite{Bergeret_RevModPhys2005,BergeretPRL,Eschrig,kadigrobov_2001}. The triplet state has three components $\uparrow \uparrow$, $\downarrow \downarrow$, $\uparrow \downarrow + \downarrow \uparrow$ that can coexist with the ordinary singlet $\uparrow \downarrow - \downarrow \uparrow$ component\cite{BergeretApplPhysA, Eschrig}. In the presence of a strong ferromagnet the opposite spins triplet component exists only near the interface. It is also the only triplet component at a clean interface when the magnetization is homogeneous\cite{BergeretApplPhysA}. This component results in a finite magnetization opposite to that of the ferromagnetic layer as has been measured at the superconducting side of a clean interfaced CoPd/Al layer\cite{Kapitulnik}. The equal spins component has been predicted to exist in the presence of spin scattering and/or in-homogenous magnetization, and is known as the long-range triplet component\cite{BergeretApplPhysA}. A supercurrent has been shown to flow through NbTi/Cr$O_{2}$/NbTi up to a distance of several hundred nanometers\cite{Keizer}, a mechanism for conversion between unpolarized and completely spin-polarized supercurrents in this case has been suggested\cite{Eschrig}. A supercurrent was also observed in a CuNi/Co/CuNi based josephson junction\cite{TruptiCoJosephson}. The CuNi, a weak ferromagnet, supplies the magnetic in-homogeneity.\\

There are two types of triplet pairings\cite{Eschrig,BergeretApplPhysA} : i) conventional triplet - which is even in frequency and odd in momentum, often called p-wave triplet. ii) odd-triplet  - which is odd in frequency and even in momentum, e.g s-wave.  Which of the two types of pairings occurs at S/F interfaces in still controversial.  It was commonly believed\cite{BergeretApplPhysA,Volkov} that in S/F hybrids the odd- triplet type occurs. However, it has also been argued\cite{Eschrig,EschrigPwave} that the even-frequency p-wave pairing is important as well.\\

Recently we have reported\cite{AlmogTripletPRB} non-local measurements on multi terminal Co/CoO/In devices consistent with the existence of an equal spin triplet order parameter (OP). However, so far there has been no direct spectroscopic evidence of a triplet OP at S/F junctions.
Here we present measurements of Co / CoO / In junctions that display two distinct features: i) a V-shaped conductance characteristic and ii) a gap larger than that of indium. These features exclude pure s-wave pairing, they point to a nodal OP. Our data is well fitted by a two fluid model incorporating a s-wave component and an equal spin p-wave component taking into account the magnetization of the Co layer. We believe that the p-wave component is generated by the spin active CoO at the Co/In interface.\\

\begin{figure*}
  \begin{center}
  \begin{minipage}[t]{1\linewidth}
    \begin{minipage}[t]{0.8\linewidth}
      \raisebox{-6cm}{\includegraphics[height=.45\textheight, width=1\linewidth,angle=0]{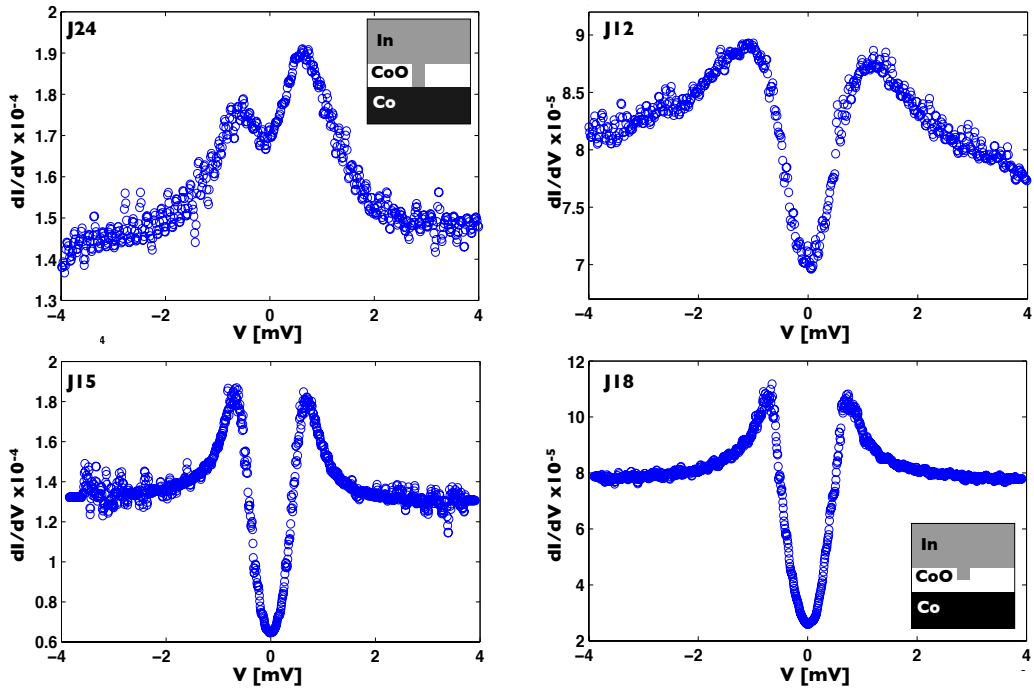}}
    \end{minipage}\hfill
    \begin{minipage}[t]{0.18\linewidth}
      \caption{Differential conductance measurements of Co / CoO / In junctions. All measurements were performed at a temperature of 0.5K.\label{fig:dIdVs}}
    \end{minipage}   
    \end{minipage}
  \end{center}
\end{figure*}

\section{experimental details}
We fabricated Co / CoO / In junctions in a cross geometry by means of e-beam lithography in a two-step process. Co bars $30 nm$ thick, $500 nm$ wide with a resistivity of $20\mu \Omega cm$ were evaporated first using e-gun evaporation at $10^{-8}$ Torr. In the second ex situ step, $60 nm$ thick, $500 nm$ wide with a resistivity of $3\mu \Omega cm$ indium bars were evaporated across using a thermal source with the substrate cooled to liquid nitrogen temperature. Between the two steps the Co surface oxidizes and develops a CoO layer $\sim 2nm$ thick~\cite{CoOThickness}, which has been shown to be antiferromagnetic and spin active~\cite{CoOxideActive1,CoOxideActive2}.\\

\section{results and discussion}
Fig. ~\ref{fig:dIdVs} shows the differential conductance $\frac{dI}{dV}(V)$ of four junctions. Measurements were taken at $0.5K$. The indium film had a $T_{c}$ of $3.5K$.  All junctions show a flat normal state differential conductance at high bias values, two conductance peaks and a reduction in differential conductance towards zero bias. Table ~\ref{table:Jpropeties} summarizes each of the junction's key features. The normal state resistance of all the junctions is several kilo-ohms. The differential conductance at zero bias compared with the normal state differential conductance shows a reduction of $70 \%$ for J18, $50 \%$ for J15, $10 \%$ for J12 and an enhancement of $30 \%$ for J24. The conductance peaks are located at a bias of $\sim0.7mV$ for J18 and J15, $1mV$ for J12 and $0.62mV$ for J24.\\
\begin{table}[!htp] \caption{Junction's main properties} 		
\centering 
\begin{tabular}{cccccc} 
\hline\hline 
ID  & Normal & Norm.  & Cond. &  Lower limit & Asymmetry \\  
   &  State &  Zero Bias & Peaks  & of $\frac{2 \Delta}{k_{B} T_{c}}$ & $10^{-6} [Siem]$ \\ 
   & Res. [K$\Omega$] & Cond. & $[mV]$ &  &  \\ [0.5ex] 
\hline  
J24 & 7.1 & 1.3 & 0.62$\pm$0.09 & 4.1$\pm$0.59 & 5\\ 
J12 & 13 & 0.9 & 1.00$\pm$0.16 & 6.6$\pm$1.05 & 2\\
J15 & 7.2 & 0.5 & 0.68$\pm$0.07 & 4.5$\pm$0.46 & 1\\
J18 & 12.5 & 0.3 & 0.70$\pm$0.07 & 4.6$\pm$0.46 & 0  \\ [1ex]    
\hline 
\end{tabular} 
\label{table:Jpropeties} 
\end{table}

In superconductor / metal junctions the barrier height determines the sub-gap conductance. A high barrier gives a conductance sensitive only to the density of states, thus the low bias conductance at low temperatures is ideally zero due to the absence of states in the superconductor. A very low barrier gives an enhanced sub-gap conductance compared with the normal state. This enhancement is due to the process of Andreev-Saint-James (ASJ) reflections\cite{Andreev,StJames,StJames2}, whereby an incoming electron from the normal metal is reflected as a hole, thus creating a Cooper pair that carries a charge of $2e$ into the superconductor. ASJ reflections are the only way to directly probe the superconducting condensate\cite{DeutscherRMP}. The four junctions presented span a wide range of barriers from a $70\%$ reduction in the zero bias conductance for J18 to an enhancement in zero bias conductance for J24. For a junction to show sub-gap conductance larger than in the normal state it must allow for ASJ reflections to dominate, namely to be sufficiently transparent. But in addition the effective contact must be smaller than the superconducting coherence length to avoid proximity\cite{BlonderPointContact} and smaller than the mean free path to avoid scattering, loss of coherence and heating at the junction. This type of contact is known as a Sharvin contact\cite{Sharvin}. A perfect one quantum channel Sharvin contact gives a resistance of $\frac{h}{e^2}=25.8K \Omega$ in the normal state of the junction. J24 has a normal state resistance of $~7.1K \Omega$ and shows a conductance enhancement which means that it probably consists of 4 quantum channels. This means it is a perfect Sharvin contact. At the other end of the barrier spectrum we have J18, a relatively high barrier tunnel junction, which mostly probes the superconducting density of states. \\

\begin{figure}
  \begin{center}
  	\includegraphics[height=.3\textheight, width=1\linewidth,angle=0]{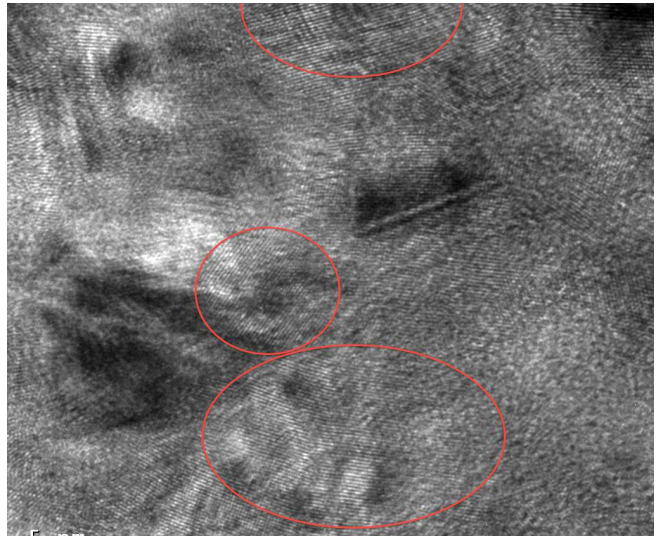}
      	\caption{Transmission electron microscope image of a 25nm thick Co film. The image shows Co grains from 5nm to 25nm. Three of the grains are outlined for clarity.)}
	 \label{fig:HiRes25nm}
  \end{center}
\end{figure}

We can estimate the effective junction area $A$ using the relation $G \sim G_{0} \frac{A}{1+Z^{2}}$, where $G_{0}$ is the quantum conductance and Z is the barrier height parameter. Since the low bias conductance (compared with the normal state conductance) of our junctions spans from a $30\%$ enhancement to a $70\%$ reduction, the Z parameter for all our junctions will be between 0.5 and 3, which gives an effective junction lateral size of several angstroms, far smaller than the actual contact size. This shows that the transport in our junctions is actually through pinholes of a few angstroms. The pinhole itself may be highly transparent (almost no oxide) and will give rise to the ASJ reflections dominated sub-gap conductance as in the characteristic of J24, this is illustrated in the inset of J24's differential conductance on Fig.~\ref{fig:dIdVs}. The pinhole may also be a region with oxide, yet substantially thinner than the surrounding oxide, this will suppress the ASJ reflections and the junction will be a tunnel junction, this is illustrated in the inset of J18's differential conductance on Fig.~\ref{fig:dIdVs}. Junctions J15 and J12 fall in between these two cases. The small pinholes here are an improvement over our previously reported work~\cite{hacohen-gourgy:152502,Almog2009} where junctions had substantially lower resistances (up to $10^{2}\Omega$ instead of $10^{4}\Omega$) and accordingly larger pinholes. The current at bias voltages just above the gap exceeds the critical velocity~\cite{hacohen-gourgy:152502} and gives rise to large dips commonly known as critical current effects. The critical current effects obscure the conductance spectra, such that it could not be fully analyzed~\cite{Almog2009}. Due to the small sized pinholes, the junctions presented here do not suffer from these effects. We have also performed transmission electron microscopy (TEM) measurements on Co films evaporated at 3 thicknesses: 6nm 12nm and 25nm. We found that the size of the Co grains spans from 5nm to the film thickness, as shown in the TEM image of the 25nm thick sample, Fig.~\ref{fig:HiRes25nm} . The highly varying grain size could be the source of the non-uniform oxide that develops on top, which in turn may be the source of the pinholes.\\

\begin{figure*}
  \begin{center}
  \begin{minipage}[t]{1\linewidth}
    \begin{minipage}[t]{0.8\linewidth}
      \raisebox{-6cm}{\includegraphics[height=.45\textheight, width=1\linewidth,angle=0]{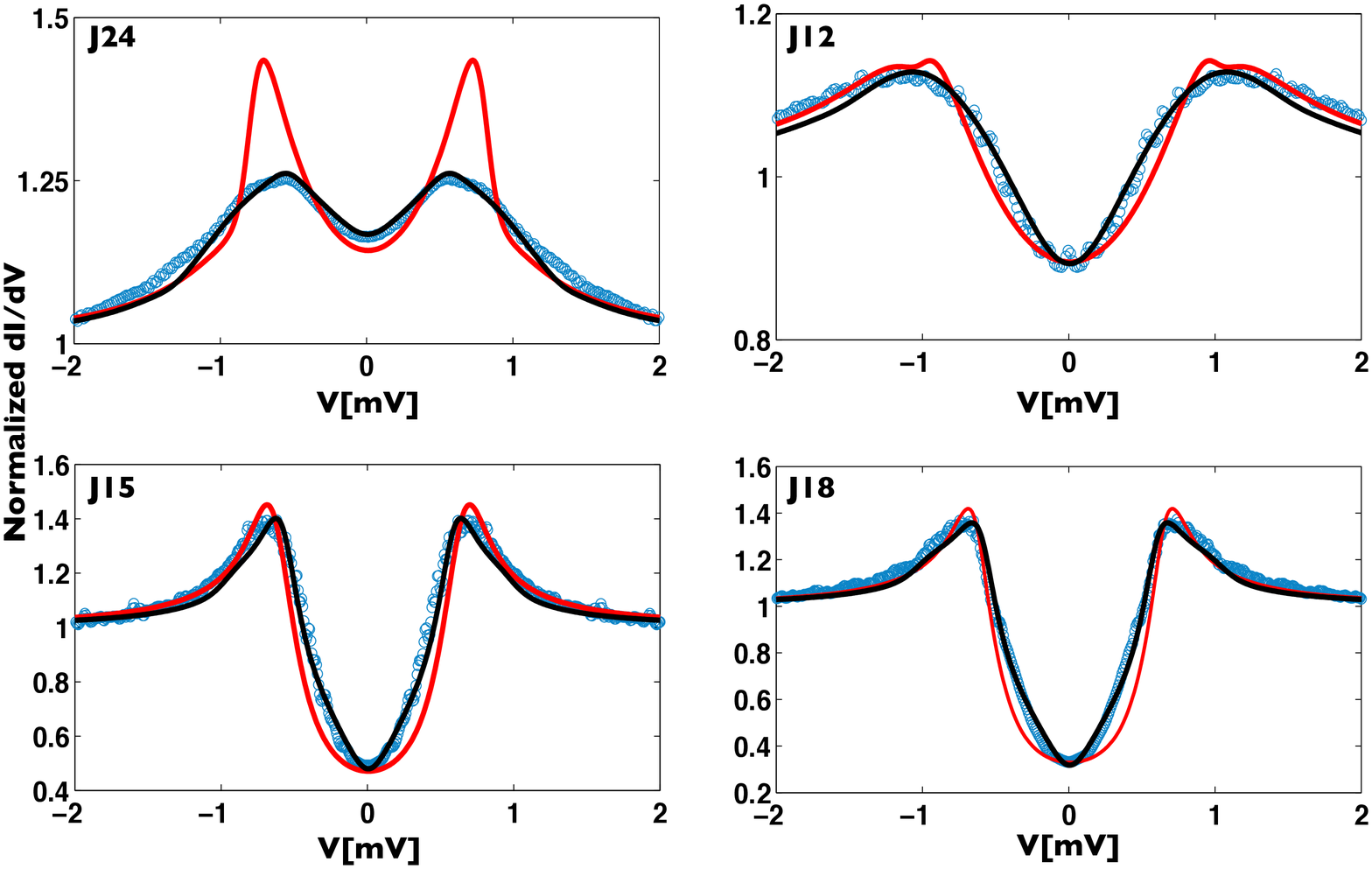}}
    \end{minipage}\hfill
    \begin{minipage}[t]{0.18\linewidth}
      \caption{Differential conductance (Symmetric component for J24 and J12) of Co / CoO / In junctions at 0.5K (blue circles). Fits to s-wave model (red line). Fits to two fluid model (black line). The fitting parameters are given in table. ~\ref{table:Jfits} \label{fig:Fits}}
    \end{minipage}   
    \end{minipage}
  \end{center}
\end{figure*}

\begin{figure}
  \begin{center}
  	\includegraphics[height=.3\textheight, width=1\linewidth,angle=0]{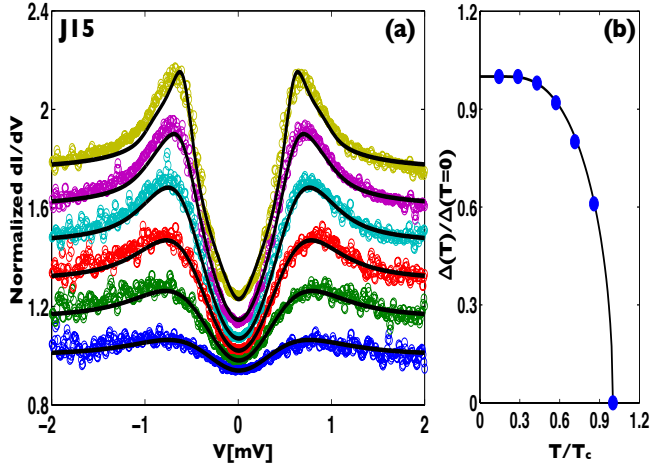}
      	\caption{(a) Temperature dependance of the two fluid fit for J15. The temperature are 0.5K to 3K at 0.5K increments. Data is shifted for clarity. (b) Gap values (relative to zero temperature gap) for each fit as function of temperature (blue circles).  BCS gap temperature dependance (black line)}
	 \label{fig:J15TwoFluids}
  \end{center}
\end{figure}

At temperatures much lower than the superconducting critical temperature $T_{c}$, the bias at the conductance peaks for a \textit{s-wave} superconductor corresponds to the superconducting gap, regardless of the barrier height. This is true whether a junction is in the Sharvin limit and gives rise to strong ASJ reflections or if it is a tunnel junction and reflects the superconducting density of states\cite{BTK_1982}. It is also true regardless of whether the normal metal is or is not magnetic\cite{mazin_2001,Perez-Willard_2004,Strijkers_2001}. Bulk indium has an energy gap value of $0.525meV$, which gives  $\frac{2 \Delta}{kT_{c}}=3.63$\cite{Giaever2DKT}, close to the weak coupling BCS limit $\frac{2 \Delta}{kT_{c}}=3.5$. The bias value of the conductance peaks of all of our junctions is at least $0.6meV$ and gives $\frac{2 \Delta}{kT_{c}}$ from 4.1 and up to 6.6. For a p-wave OP the conductance peaks are located at a bias value \textit{lower} than the maximum gap\cite{Tanaka,Laube}. In this case the bias at the conductance peaks gives only a lower limit on the gap value. This lower limit experimental value for the known triplet p-wave superconductor $Sr_{2}RuO_{4}$ is $\sim 6.8$\cite{Laube}. Combescot\cite{Combescot} calculated the $\frac{2 \Delta_{M}}{kT_{c}}$ ratio in the weak coupling limit for highly anisotropic OPs and obtained values between 4 and 4.5. For d-wave and p-wave the results are $\frac{2 \Delta_{M}}{kT_{c}}=4.26$, yet at strong couplings the values can be as high as 10\cite{Scalapino2DKT}. One may claim that spreading resistance may cause the large values observed, but from the resistivity values we can see that the electrode resistance ($R_{\square}$) is at least three orders of magnitude lower than the junction resistances which rules out any spreading resistance effects.\\

We now analyze in detail the structure of the differential conductance. Before we do so we note that our junctions show a slight asymmetry with respect to the bias voltage. The asymmetry is seen also above $T_{c}$, hence it is a property of the normal state. We define it as the difference between the positive high bias and negative high bias values of the differential conductance, the values are given in table ~\ref{table:Jpropeties}. As the barrier is increased the asymmetry diminishes and is negligible for the two high barrier junctions.\\ There is at least one paper\cite{TripletAsymm} that assigns an asymmetry in transport to a triplet superconductor, but only in the sub-gap bias region. When comparing the data to models we eliminated the asymmetry of the two low barrier junctions by taking the symmetric part of the data only. \\

In our junctions the differential conductance exhibits more of a V-shape structure than a conventional S/F junction\cite{Perez-Willard_2004}. Nevertheless we have tried to fit our data with the conventional S/F model of Perez-Willard \textit{et. al.} \cite{Perez-Willard_2004}, who have shown a very good agreement between their $5nm$ porous Al / Co junction characteristics and their S/F s-wave singlet model. Our junctions do not fit this model, or any other model\cite{Strijkers_2001,mazin_2001}, which predicts a U-shape for the characteristic conductance up to the gap. To demonstrate this we show best fits to the data, red curves in Fig.~\ref{fig:Fits}. They show two distinct differences with the data. First, the sub gap conductance is V-shape rather than U-shape, as already said. Second, the conductance peaks are wide rather than sharp and narrow. Moreover, as can be seen in table~\ref{table:Jfits}, some of the parameter values are not so reasonable, physically. The gap values given by the fits are $14\%$ to $90\%$ higher than indium. Also the polarization is lower than the measured value for Co using similar techniques, $P\sim42\%$\cite{Perez-Willard_2004,Soulen}.\\

\begin{table}[!htp] \caption{Fitting parameters.\\ s-wave model parameters: $\Delta [meV]$ - superconducting gap, $t_{\uparrow,\downarrow}$ - transmission probabilities for spin $\uparrow / \downarrow$, P - ferromagnet's polarization\\ triplet model parameters: $\Delta_{\uparrow , \downarrow}$ - gap for each spin, $Z$ - barrier height, $\Gamma_{S,P}$ - quasiparticle life time broadening relative to the gap, R - $\frac{s-wave}{p-wave}$ fluid ratio} 		
\centering 
\begin{tabular}{ccccc} 
\hline\hline 
Parameter  & J24 & J12  & J15 & J18 \\ [0.5ex] 
\hline 
s-wave model\\
\hline 
$\Delta [meV]$ & 0.8 & 1.0  & 0.63 & 0.6\\ 
$t_{\uparrow}$ & 0.97 & 0.93 & 0.66 & 0.57  \\
$P=\frac{t_{\uparrow}-t_{\downarrow}}{t_{\uparrow}+t_{\downarrow}}$ & 0.25 & 0.28 & 0.25 & 0.30 \\
$\Gamma$ & 0.30 & 0.37 & 0.10 &0.10 \\
\hline  
two fluid model\\
\hline  
R & 0.25 & 0 & 1 & 1 \\ 
$\Delta_{\uparrow,\downarrow} [meV]$ & 1.32 & 1.50 & 0.95 & 1.00 \\ 
$Z$ & 0.70 & 1.23 & 1.85 & 2.50 \\
$\Gamma_{P}$ & 0.04 & 0.12 & 0.00 & 0.00 \\ 
$\Delta_{S} [meV]$ & 0.58 & - & 0.59 & 0.57 \\ 
$t_{\uparrow}$ & 0.986 & - & 0.720 & 0.540 \\ 
$\Gamma_{S} $ & 0.17 & - & 0.04 & 0.07 \\ 
\hline 
\end{tabular} 
\label{table:Jfits} 
\end{table}

To fit the data we have assumed the coexistence of a s-wave and a triplet p-wave fluid. To represent the s-wave we stick with the model by Perez-Willard \textit{et. al.} \cite{Perez-Willard_2004}. We have added to it the conductance of a triplet equal spin p-wave / ferromagnet, as calculated by Linder \textit{et. al.} \cite{PhysRevB.75.054518}. The fits show a very good agreement with the data. The low temperature fits and the fitting parameters are shown in Fig.~\ref{fig:Fits} and table \ref{table:Jfits}. We also present the temperature dependence for J15 in Fig.~\ref{fig:J15TwoFluids}. All the parameters were held at the low temperature values, except for the p-wave and s-wave gaps. Their ratio (R), was assumed to remain the same as at 0.5K. The panel of Fig.~\ref{fig:J15TwoFluids}.b shows that the gaps follow the BCS gap temperature dependence. For all fits the Co polarization was taken as 0.42 like previously measured values\cite{Perez-Willard_2004,Soulen}.  Transport is assumed to be in the node direction and the superconductor's polarization is zero, yet raising it a little hardly changes the fit quality, up to about $6\%$. We may also note that the $\Gamma$ parameter values needed to obtain the fits are substantially lower in the two fluid model compared with the best fits of the s-wave model. This further supports the two fluid model.\\

The triplet component is induced through ASJ reflections and these are suppressed in the tunneling limit, thus we expect that the s-wave component will be largest for the higher barrier junctions. The $\frac{s-wave}{p-wave}$ fluid ratio for the two high barrier junctions J15 and J18 is 1, whereas J12 could be fitted without any s-wave fluid and J24 only needed a small fraction. As one might expect, the two low barrier junctions show a larger p-wave gap.\\

Since the $T_{c}$ of our thin indium film is $3\%$ larger than bulk indium, the s-wave gap is also expected to be larger, we find it to be up to $10\%$ larger. Also the data suggests that the nodes are aligned perpendicular to the junction. We are unaware of any predictions regarding the direction.\\

\section{summary}
The differential conductance of Co / CoO / In contacts exhibits, at low bias, a V-shape structure that is incompatible with a purely s-wave symmetry OP. The bias at the conductance peaks is larger than the gap of indium for all contacts, which also rules out s-wave symmetry. A two fluid model comprising of an equal spin triplet p-wave and singlet s-wave fluids allows to successfully fit the shape of the entire conductance characteristic. We find that the values of the s-wave and p-wave gaps follow the BCS temperature dependance. The model applies equally well to weak and strong barrier contacts with the fits giving a consistent set of parameters including the ratio of the two fluid amplitudes and the size of the p-wave gap. We believe that the thin CoO, being spin active~\cite{CoOxideActive1,CoOxideActive2, CoOThickness}, provides sufficiently strong spin scattering, which creates the triplet OP component.\\
This work was supported in part by the Israel National Science Foundation (Grant No. 481/07).



\bibliography{mySFbib}

\end{document}